# Optimizing the flux coupling between a nanoSQUID and a magnetic particle using atomic force microscope nanolithography

M. Faucher[1,*], P.O. Jubert[1,+], O. Fruchart[1], W. Wernsdorfer[1] and V. Bouchiat[1]

[1]*Institut Néel, CNRS & Université J. Fourier, BP 166, 38042 Grenoble Cedex 9, France*

*now at IEMN-Lille, Villeneuve d'Ascq, France.

+ *Now at* IBM-Almaden, CA, USA .

We present results of Niobium based SQUID magnetometers for which the weak-links are engineered by the local oxidation of thin films using an Atomic Force Microscope (AFM). Firstly, we show that this technique allows the creation of variable thickness bridges with 10 nm lateral resolution. Precise control of the weak-link milling is offered by the possibility to real-time monitor weak-link conductance. Such a process is shown to enhance the magnetic field modulation hence the sensitivity of the magnetometer. Secondly, AFM lithography is used to provide a precise alignment of NanoSQUID weak-links with respect to a ferromagnetic iron dot. The magnetization switching of the near-field coupled particle is studied as a function of the applied magnetic field direction.

1) **Introduction**

Atomic Force Microscope (AFM) lithography based on nano-oxidation of metallic thin films is a unique technique for direct patterning of electronic devices at the nanometer scale [1]. The process involves the local surface oxidation of the metallic film under ambient conditions with a voltage-biased tip. Applied on ultrathin films, this allows the patterning of electrically-insulating oxide patterns (Fig.1a) of typically 10nm nanometer linewidth. It has been used by Matsumoto et al. [2] for the realization of nanoscale single electron devices showing room temperature operation [3, 4].

More recently the same technique has been applied to realize superconducting nanostructures [5] based on nanoscale constrictions, allowing the fabrication of niobium [5] and niobium nitride [6] nanoSQUIDs as well as single photon detectors [7]. Depending on the operating conditions (tip voltage bias and speed) and film thickness, oxidation of the film can be performed through part or all of the thickness (see Fig. 1b) of the entire film. The amount of metal transformed into the oxide being can be indirectly determined by the measurement of the oxide protrusion. This interesting feature allows a three dimensional control of a superconducting nanostructure, similar to what can be obtained using focused ion beam (FIB) milling [8-10]. Indeed AFM local oxidation has comparable resolution as the one obtained using FIB milling. It also benefits from some other advantages such as the absence of contamination and the possibility to protects the nanostructure against further oxidation since the resulting nanostructure is embedded in the oxide generated by the anodization. Note however that unlike a FIB beam which can penetrate thick samples, AFM oxidation is restricted to ultra-thin films as the oxidation can hardly be performed on thickness exceeding 10nm. In this paper, we illustrate the versatility of this technique by extending its use on more complex structures. We are describing here two different experiments, which use the AFM oxidation lithography to engineer the superconducting weak-link and to optimize its operation. The first experiment involves the post-processing of a SQUID made by electron–beam lithography (EBL) while the second is dealing with the full fabrication of a nanoSQUID within multilayers made of ferromagnetic dots coated with superconducting niobium.

## 2) In Situ milling of Niobium NanoSQUID junction using AFM oxidation

We first describe the AFM-based post-processing of micro-SQUID devices made by AFM oxidation on SQUID devices made using EBL and plasma etching [11]. EBL allows the processing of a large number of devices with homogeneous geometrical factors, while AFM lithography allows one to refine the working characteristics of the bridges. On the tested sample, the two SQUID weak-links are initially made of a lateral constriction of constant thicknes (following the so-called "Dayem" bridge [12] geometry). The SQUID is first imaged with the AFM without tip bias to precisely align the tip over the Dayem Bridge. Their original size prior to AFM lithography consists of a nanowire of 60 nm width and 250 nm length. The AFM oxidation is then performed by proceeding to a single line scan perpendicular to the Dayem bridges (white arrow in inset of Fig.2) applying a voltage of about -10V on the tip, with a relative humidity exceeding 50%. The effect of the oxidation can be monitored either directly by measuring in real-time the nano-bridge resistance increase (the two terminal resistance increases by typically a few percents) or by imaging the oxide protrusion during a subsequent AFM scan without applied voltage (see Fig.2). Indeed, the oxidation process results in a swelling of the bridge under the tip trajectory. Under repeated scan using low scan speeds (~0.1µm/s), the oxidation is homogeneously done below the tip trajectory resulting in an even swelling (Fig. 2 top). If the tip scan speed is increased in the range of 1 – 3 µm/s, the oxidation only occurs on the bridge edges as evidenced by the observation of the oxide protrusion of oxide only on the bridge sides (see Fig. 2, bottom). This selectivity effect is attributed to the stronger interaction of the wire edge with the biased AFM tip. In order to illustrate the power of AFM oxidation, we selected a sample which exhibits a rather poor magnetic flux modulation of switching current $I_{sw}$ before processing (black curve of Fig. 3). We have tested the effect of the oxidation procedure by measuring the magnetic field modulation at low temperature of the same SQUID before and after the oxidation step (Fig. 3, black and blue curves, respectively). The first observation is that the maximum switching current of the trimmed SQUID is reduced by roughly a factor 2 due to the decreased cross-section of the weak links. However it exhibits a much better magnetic flux (Fig. 3) dependence with a typical saw-tooth-shaped modulation. This last effect can be understood as a consequence of the change in the local topology in the weak-links resulting from the oxidation. since the SQUID interference contrast in a SQUID with "long" Dayem bridges (i.e. which bridge lengths are longer than the superconducting coherence length) is known to be inversely proportional to the kinetic inductance [13] of the bridge, the latter being directly proportional to $L/S$ [14], where $L$ and $S$ are respectively the length and the cross-section of the weak-link [6]. The effect of the oxidation is to locally decrease $S$ (say by a factor 2) on a lateral size $L'$ of the order of the lateral size of the patterned oxide line (around 10 nm) [5], which is much smaller than $L$. One then transforms the effective weak-link from the class of constant thickness ("Dayem") bridges [12] towards the class of "variable thickness bridges (VTB)", [13] . Furthermore in VTB based SQUID, the torsion of the phase induced by the circulating current through the weak link is more localized compared to devices with Dayem bridges weak links [15] Note that at magnetic field values $|B| \gtrsim 25\, \Phi_0$, the current of the VTB SQUID exceeds the Dayem SQUID and the two modulation curves intersect with each other. Indeed a smaller weak link is less sensitive to the penetration of the magnetic field. Therefore, the shrinking the weak link size results in widening the envelope of the modulation curve (known as Fraunhofer pattern). At higher fields where the SQUID is no more periodically modulating, the switching current exhibits similar variations before and after AFM oxidation: for example, the variations recorded between -60 and -40 $\Phi_0$ show clear correlation . On the other side (i.e. 30 to 60 $\Phi_0$ ) such similar correlation can still be seen but the VTB SQUID show remnant periodic oscillations superimposed on that signal.  The un-modulated component presumably comes from depinning of vortices in the superconducting stripes away from the weak-links. Since these parts are not affected by the AFM lithography step, their behaviors are unchanged by the weak-link engineering.

### 3) AFM made nanoSQUIDs coupled to a single self-assembled nanoscale iron dot

The second experiment illustrates the potential of AFM nano-oxidation as a stand-alone fabrication technique. For that purpose, we use the oxidation process on a plain multilayer substrate to pattern a nanoSQUID as a final fabrication step. This order allows us to accurately position the detector with respect to the probed magnetic object thus optimizing flux pick-up.

The chosen substrate is a superconducting/ferromagnetic multilayer where a nanoSQUID junction must be coupled to a single nano-magnet. Indeed it has been shown [16],[17] that the best inductive coupling of a nanomagnet to a nanoSQUID is achieved when the magnet is in direct proximity to one of the junctions.  The magnet chosen for this experiment is a sub-micrometer sized faceted iron dot, which exhibits high structural quality and display simple magnetic flux-closure states. These  dots are self assembled and then encapsulated under a protective capping layer in an ultra-high vacuum using pulsed-laser epitaxy [18]. They are grown on a sapphire substrate of orientation $(11\bar{2}0)$ initially buffered in situ with a 8-nm-thick molybdenum film, which provides the necessary conditions to give rise to self-assembly. Note that besides the predominant iron dot growth, there exists another sort of bi-dimensional nanostructures attributed to iron-molybdenum surface alloy (which can be seen in Fig. 4a as thin, randomly-shaped mesa labeled "Fe-2D" ), partially wetting the molybdenum film. A capping layer made of a trilayer of oxidized aluminum (2 nm) followed by niobium (15 nm) and silicon (2 nm) is evaporated. Each film of the capping layer has a given purpose. The niobium layer provides the source of superconductivity and is protected from contamination and oxidation by both the under and over layers. The aluminum oxide buffer layer intercalated between the iron particle and the niobium is effective to reduce the inverse proximity effect that plagues superconducting thin films in close contact with ferromagnetic materials.

Since the dot formation results from a self-assembly process, its position on the film is not controlled. This implies that the nanoSQUID junction has to be precisely aligned with respect to the magnetic particle to be measured. The easiness of alignment provided by scanning probe microscopies offers a good opportunity to precisely align a given nanostructure with respect to another nano-object. In previous experiments, [19], alignment was obtained statistically by selecting "good" devices among a large ensemble of batch-processed samples. The present technique offers the advantage to allow working on prototypes existing as single samples only. Alignment of the AFM tip prior to oxidation patterning step can be performed with an extremely good accuracy (precision better than 10 nm) if one uses an Atomic Force Microscope equipped with piezoelectric actuators corrected for their non-linearity (so-called 'closed-loop scanner'). One should note the film is not altered during this alignment procedure since imaging is performed in absence of voltage applied on the tip.

We have selected a Fe dot of size 200 x 50 x 30 nm (Fig 5a). Both SQUID loop and lateral constrictions are made of fully oxidized patterned lines using conditions described in [5]. The low density allows the study of a single dot coupled to a nanoSQUID magnetometer. The resulting structure is presented in Fig. 4, bottom. It consists of a circular oxide circle and two lateral oxide wedges defining 40 nm wide Dayem constrictions. A minimal distance of 400 nm is maintained between the magnet and the oxide loop in order to preserve the ferromagnetic particle from the oxygen diffusion induced by the AFM lithography (Fig 5, a). The sample is then transferred in a dilution fridge equipped with 3D superconducting coils which allow the study of magnetization reversal properties below 40 mK similarly as performed in [20-22]

Typical distances between dots are such that the magnetostatic interactions between them can be neglected, so that they behave essentially like isolated nanomagnets. These have been shown experimentally by Magnetic Force Microscopy and by micromagnetic simulation to display flux closure magnetic domains [23]. Such property can be understood as Fe is a rather soft magnetic material. Micromagnetic simulations later predicted the magnetization processes under field of such dots: starting from saturation one or several magnetic vortices or domain walls enter the dot at its edges upon reducing the external field (nucleation events), yielding the flux-closure state at remanence. Upon increasing again the external field the magnetic vortices and domain walls are expelled at an edge (annihilation events). It is the fields at which the nucleation or annihilation events occur, that the microSQUID will detect in the so-called "cold mode" [14,21]. The magnetization process of these dots has been simulated in details in and microSQUID measurements are been obtained in [19] confirming a multi-domain structure with clear angular dependence of the magnetization. This magnetization depends on both intrinsic (shape, anisotropy) and extrinsic (microstructure, defects) parameters. The dots (see AFM micrograph in Fig. 4a, inset) have an elongated hexagonal shape and exhibit atomically flat facets. The spacing between dots (about 1 µm in our films, see Fig. 4a) is such that inter-dots dipolar interactions are negligible. The magnetic field is applied within the plane of the SQUID and ramped from zero to a maximum amplitude of about 0.3 T at a given and fixed azimuth angle. This azimuth angle is varied scan after scan in given step in order to cover the whole *x-y*-plane, while a feedback is applied in the *z*-direction to keep the SQUID at the same working point (i.e. a constant switching current). Flux jumps are recorded and analyzed and then plotted in the *x-y*-plane. Fig. 5, b shows the plots of the switching values of the nanoSQUID recorded at in the *cold-mode* method [14,21] here performed at 40 mK,. The jumps may come from various origins: the sought signal, i.e. jumps associated to nucleation/annihilation events in the 3D Fe dots, and parasitic signals including the magnetization reversal of the 2D Fe platelets, or the motion of unpinned vortices within the Nb. In this plot, one can highlight four colored branches exhibiting a point symmetry with respect to the origin, whose shapes are portions of an 3D geometrical astroid [24]. Such individual shape is similar to the model proposed by Stoner and Wohlfarth [25] to describe a single domain magnetization reversal. Entangled astroids have already been observed in another microSQUID experiments where the microSQUID had been patterned with a regular EBL technology. This feature is expected in the case of flux-closure dots were several nucleation/annihilation events may be expected to turn a single-domain state into a flux-closure state, and vice-versa [19].

**4) Conclusion**

These two series of experiments bring insights of the potential of scanning probe nanolithography for optimizing the operation of a NanoSQUID interferometer. The easiness and precision of alignment and the possibility of an easy in-situ control of the device geometrical and electrical parameters are unique features associated to this technique. We have shown that this lithography allow to increase magnetometry sensitivity the near field regime by optimizing two independent parameters . Firstly the miniaturization of the weak links increase the magnetometer responsivity, Secondly the easiness of the alignment offers new possibilities to increase the near-field inductive coupling between the magnetometer and the magnetic elements to be studied.

**Acknowledgments**

D. Mailly is gratefully acknowledged for providing us the SQUID made by electron-beam lithography on which AFM nano-oxidation has been performed.

**Figures :**

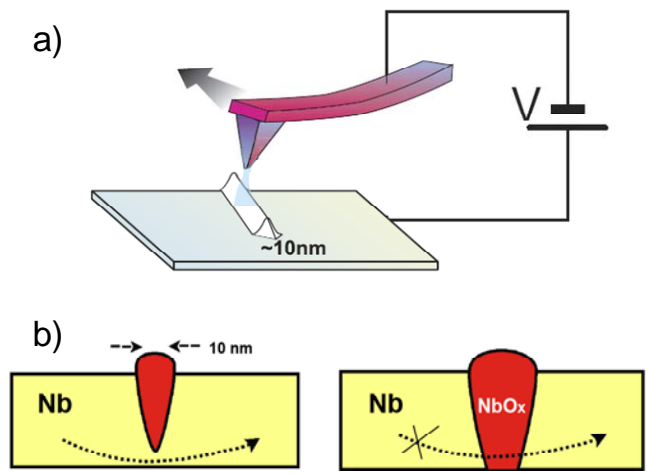

**Figure 1.** a): Principle of the AFM nano-lithography by oxidation,
b) Cross section of an AFM made oxide line drawn on a metallic ultrathin film. Depending on the oxidation parameters (tip bias, speed, humidity), oxidation of the film can be obtained in the partial or full thickness of the film.

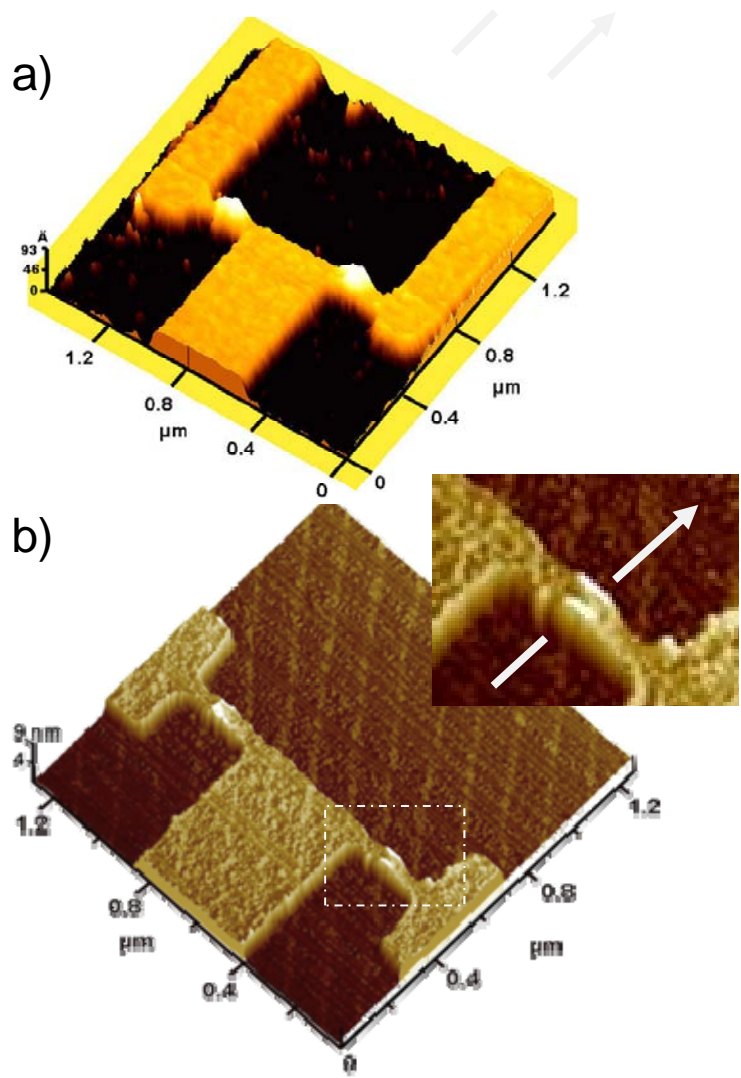

**Figure 2.** AFM micrographs of SQUID weak links made of Dayem bridges partially oxidized using AFM lithography. a) using a low tip scan speed (~0.1µm/s) : the weak link section is homogeneously reduced as shown by a regular swelling of the nanowire. (b) using a high scan speed of the tip : only the edges of the weak links are oxidized (see zoom in inset). The white arrow indicates the tip motion during lithography.

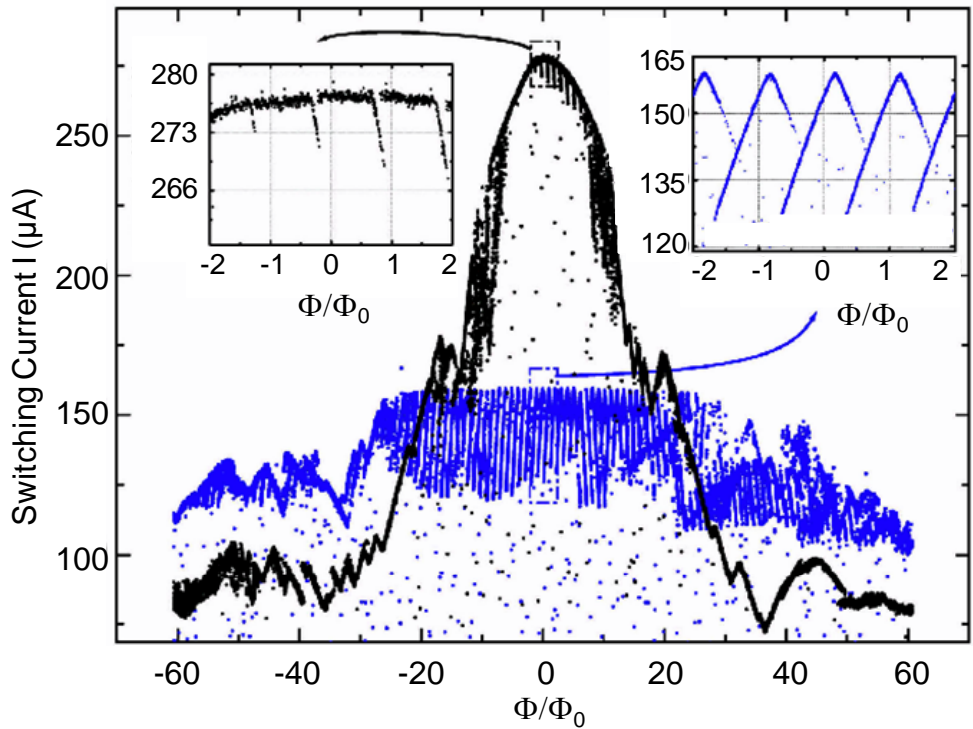

**Figure 3.** Measurement of the switching current at 30mK of the NanoSQUID as a function of the applied magnetic field (shown in reduced units of flux quanta ) before (black dots) and after (blue dots) AFM oxidation of the nanobridges.

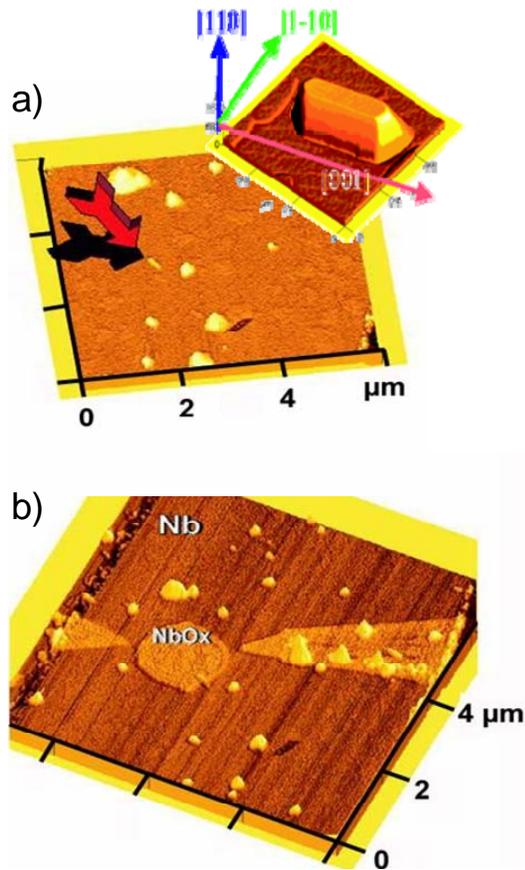

**Figure 4.** (a): AFM Micrograph of the multilayered sample before the lithography step. The arrow indicated the ferromagnetic particle chosen to be coupled to the NanoSQUID magnetometer. Inset : Zoomed AFM micrograph of an Iron magnet showing the faceted structure of the submicron ingot. (b) AFM micrograph of the same sample after the AFM patterning of the oxide nanostructures. It is composed of three features : a central disc which act as the SQUID loop and two lateral wedges. The two wedge-disk gaps define the Dayem constrictions which will be the sensing parts of the magnetometer.

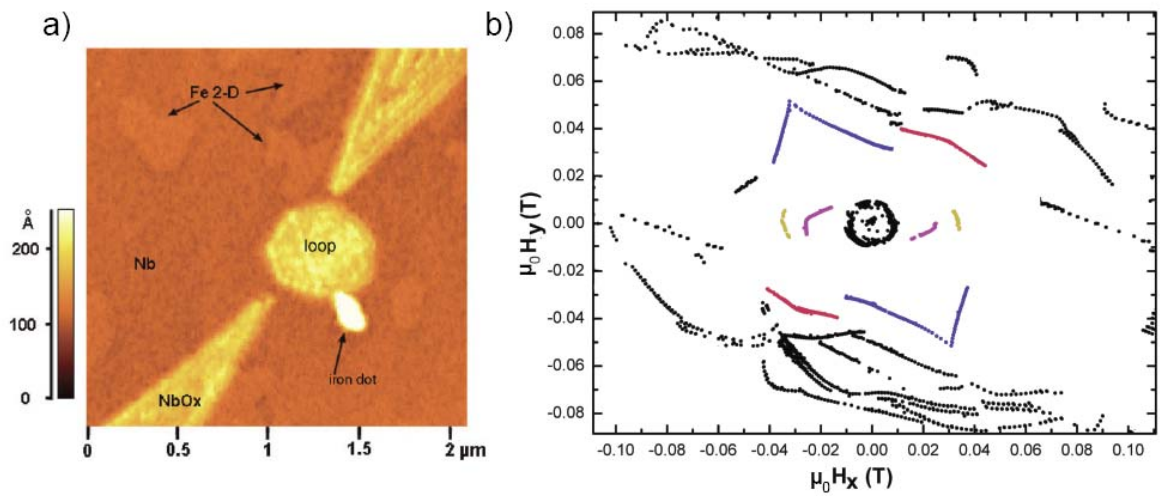

**Figure 5.** a) AFM Micrograph of the test sample showing the alignment of the patterned SQUID with respect to a single Iron dot surrounded by the patterned SQUID. b) Polar map of the switching field measured for the AFM NanoSQUID depicted in a), as a function of the Field amplitude ($H_x, H_y$). Colored lines emphasize the portions of Stoner-Wohlfarth astroids, each of them being the signature of magnetization switching events.